\documentclass[fleqn,usenatbib]{mnras}

\usepackage{newtxtext,newtxmath}

\usepackage[T1]{fontenc}
\usepackage{ae,aecompl}

\usepackage{graphicx}	
\usepackage{amsmath}	
\usepackage{amssymb}	

\title[Search for companions around $\alpha$ Centauri]{Searching for Faint Comoving Companions to the $\alpha$\,Centauri system in the
VVV Survey Infrared Images\thanks{Based on observations taken  within  the  ESO  VISTA  Public Survey  VVV,  Programme  ID 179.B-2002}}

\author[J. C. Beam\'in]{
J. C. Beam\'in,$^{1,2}$\thanks{E-mail: juancarlos.beamin@ifa.uv.cl}
D. Minniti,$^{3,2,4}$
J. B. Pullen,$^{3}$
V. D. Ivanov,$^{5,6}$
E. Bendek,$^{7}$
\newauthor
A. Bayo,$^{1}$
M. Gromadzki,$^{8}$
R. Kurtev,$^{1,2}$
P. W. Lucas, $^{9}$
R. P. Butler, $^{10}$
\\
$^{1}$Instituto de F\'isica y Astronom\'ia, Facultad de Ciencias, Universidad de Valpara\'iso, Ave. Gran Breta\~na 1111, Playa Ancha, Valpara\'iso, Chile.\\
$^{2}$Millennium Institute of Astrophysics, Chile.\\
$^{3}$Facultad de Ciencias Exactas, Universidad Andres Bello, Fernandez Concha 700, Las Condes, Santiago, Chile\\
$^{4}$Vatican Observatory, V00120 Vatican City State, Italy\\
$^{5}$European Southern Observatory, Karl-Schwarzschild-Str. 2, D-85748 Garching bei M\"unchen, Germany\\
$^{6}$European Southern Observatory, Ave. Alonso de Cordoba 3107, Casilla 19001, Santiago, Chile.\\
$^{7}$ NASA AMES, California, USA.\\
$^{8}$Warsaw University Astronomical Observatory, Al. Ujazdowskie 4, 00-478 Warszawa, Poland\\
$^{9}$Centre for Astrophysics Research, University of Hertfordshire, Hatfield AL10 9AB, UK\\
$^{10}$Department of Terrestrial Magnetism, Carnegie Institution of Washington, 5241 Broad Branch Road, NW, Washington, DC 20015-1305, USA\\
}

\date{Accepted XXX. Received YYY; in original form ZZZ}

\pubyear{2017}

\begin{document}
\label{firstpage}
\pagerange{\pageref{firstpage}--\pageref{lastpage}}
\maketitle

\begin{abstract}
The VVV survey has observed the southern disk of the Milky Way in the near
infrared, covering 240\,deg$^{2}$ in the $ZYJHK_S$ filters. We search the VVV Survey
images in a $\sim$19\,deg$^{2}$ field around $\alpha$ Centauri, the nearest stellar system to the Sun, to look for possible overlooked companions that the baseline in time of VVV would be able to uncover.
The photometric depth of our search reaches $Y\sim$19.3 mag, $J\sim$19 mag, and $K_S\sim$17 mag. This search has yielded no new companions in $\alpha$ Centauri system, setting an upper mass limit for any unseen companion well into the brown dwarf/planetary mass regime. 
The apparent magnitude limits were turned into effective temperature limits, and the presence of companion objects with effective temperatures warmer than 325\,K can be ruled out using different state-of-the-art atmospheric models.

These limits were transformed into mass limits using evolutionary models, companions with masses above 11 M$_{Jup}$ were discarded, 
extending the constraints recently provided in the literature up to projected distances of d\,<\,7\,000\,AU from $\alpha$ Cen AB and $\sim$1\,200 AU from Proxima. 
In the next few years, the  VVV extended survey (VVVX) will allow to extend the search and place similar limits  on brown dwarfs/planetary companions to $\alpha$ Cen AB for separations up to 20\,000AU.
\end{abstract}




\section{Introduction}

The nearest stellar system $\alpha$ Centauri (including the
close binary $\alpha$ Cen AB and Proxima) allows 
us to probe to unprecedented depth the vicinity of three
stars for planets. 
$\alpha$ Cen AB is three times closer than any other FGK star offering unique conditions for detection and characterization of earth-like planets around sun-like stars in terms of brightness and angular separation of a hypothetical habitable planet. However, the system has not been considered in the target list of exoplanet imaging missions because of light contamination of the environs of each binary component by the other.
Recent advances in binary star light suppression and wavefront control \citep{Thomas2015} has enabled the creation of dark zones around binary systems. As a result, dedicated mission concepts to observe $\alpha$ Centauri has been proposed \citep{Bendek2015} with telescopes as small as 40cm in aperture. Scientists and engineers \citep{Sirbu2017}
are also studying whether the WFIRST coronagraph would be able to observe binaries and include $\alpha$ Cen in the target list.

It is worth mentioning that detecting an earth-like planet in the habitable zone (HZ) of $\alpha$ Cen AB with a 40\,cm aperture telescope is equivalent, in terms of photon flux and angular separation, to performing the same detection around a star at 10\,pc with a 4-m class NASA's ``HABEX'' flagship exoplanet mission.   

This system will be an important target to be further explored  with the next generation of space telescopes and missions and of the Breakthrough Starshot 
project\footnote{\url{http://breakthroughinitiatives.org/Initiative/3}}.

A planet on a 3.2 day orbit was reported to exist around $\alpha$ Cen B \citet{Dumusque2012} but more recent studies cast doubts on the existence of this planet, arguing that the signal reported by \citet{Dumusque2012} ``arise from the window function of the observed data''\citep{Rajpaul2016}.
\citet{Demory2015} looked for evidence of $\alpha$ Cen Bb using HST/STIS photometry. They found no evidence of the proposed $\alpha$ Cen Bb, but on the other hand reported the presence of a transit like feature in the light curve of $\alpha$ Cen B, that might be produced by a earth  mass planet in a $\sim$ 15-20 day orbit. 
 \citet{Kervella2006a} studied $\alpha$ Cen AB, with the NACO instrument at VLT, and set upper limits for a possible co-moving companion in the $H$ and  $K$ band, corresponding to $\sim$20-30 M$_{Jup}$ with separations between 7 and 20 AU. Later, \citet{Kervella2007a} using optical imaging  ($V$, $R$, $I$ and  $Z$ bands) complemented this search and determined that there were no co-moving companions to this system with masses $\gtrsim$15-30 M$_{Jup}$ at separations between 100-300 AU

\citet{Quarles2016} investigated numerically if stable planetary orbits exists around one of the stars or around the $\alpha$ Cen AB binary, and arrived at a positive answer (see their Fig.\,11).

Recently, \citet{Pourbaix2016} and \citet{Kervella2016a} performed a detailed astrometric study of the $\alpha$ Cen AB system, and derived not only precise proper motion and parallaxes, but also orbital parameters and in the latter case predictions of microlensing events in the following years, that would allow to probe  an unexplored parameter space for the presence of exoplanets around those stars.

\bigskip

Regarding Proxima: \citet{Benedict1999} found no companions with masses above 0.8$M_{Jup}$ in the period range 1$\leq$ P [days] $\leq$1000, using HST Guide Sensor data. \cite{Endl2008}, based on nearly 7 years of radial velocity measurements found no evidence of planets with masses larger than  M\,sin\,(i)$\geqslant$ 1 M$_{Neptune}$, at periods $\leqslant$ 2.7 yr. 

\cite{Lurie2014} using ground based astrometric measurements constrained the presence of planets with masses down to 2\,M$_{Jup}$ with periods 2$\leq P$\,[years]$\leq$5, and down to 1$M_{Jup}$ for 5$\leq P$\,[years]$\leq$12. As pointed by \cite{Lurie2014}, these studies eliminate the possibility of finding any Jupiter like planet around Proxima for orbital periods out to 12 years.
Recently a rocky planet in a 11 day orbit was reported by \citet{anglada-escude2016} making Proxima b the closest exoplanet known. 

\cite{Mesa2017} searched for the presence of  giant exoplanets around Proxima using high contrast imaging with SPHERE instrument at the Very Large Telescope, no objects were found with masses above 6-7 M$_{Jup}$ at 0.5-1 AU, and 4 M$_{Jup}$ at distances larger than 2.5 AU, using the AMES-COND models \citep{Baraffe2003}.

We have investigated the presence of substellar companions with separations up to 7\,000 AU from $\alpha$ Cen and up to 1\,200 AU from Proxima. This paper is organized as follows: Section 2 describes the observations, section 3, the manual and automated search for faint companions. Section 4 is dedicated to discussion of the limits imposed by our search and the final section 5 gives the conclusions.

\section{Sample selection and observations}
\label{Sec:data}

\subsection{VISTA/VIRCAM}
The VVV survey \citep{Minniti2010, Saito2012, Hempel2014} was one of the six ESO public surveys carried out with the 4.1m Visual and Infrared Survey Telescope for Astronomy (VISTA) telescope and VIRCAM camera \citep{Dalton2006,Emerson2010} at cerro Paranal Chile. The VIRCAM detector has sixteen chips of 2048$\times$2048 pixels
with a pixel scale of 0.34\arcsec\ . The total area covered after six overlapping pointings (known as ``pawprints'') is 1$\times$1.5 deg (hereafter a ``tile''). VVV had a multicolour campaign in 5 NIR bands ($ZYJHK_S$) the first year (2010), and then 5 years of monitoring in the $K_S$ band, where the total number of epochs differed from field to field from almost 300 epochs in some bulge fields to 54 epochs in the least observed disk field. Additionally, during the last year of the survey  (2015) one/two extra epochs were obtained in the $ZYJH$ bands. The main goal of VVV was to trace the 3-D structure of the Milky Way, mainly through the study of variable stars \citep{Dekany2013} but also using Red clump stars \citep{Gonzalez2011} and NIR mutiwavelength  studies \citep{Minniti2014}. This survey is useful for accurate measurements of proper motions (PMs) and parallaxes, as demonstrated previously by \citet{Beamin2013,Ivanov2013,Beamin2015,Smith2015,Kurtev2017,Beamin2017,Smith2017}. 
The data used in this study, were reduced at the Cambridge Astronomy Survey Unit (CASU) with pipeline v1.3.  
In this study we considered 13 different tiles, two epochs in the $Y$ and $J$ bands and three epochs for the $K_s$ band.

\begin{figure}
\centering
\includegraphics[width=\linewidth]{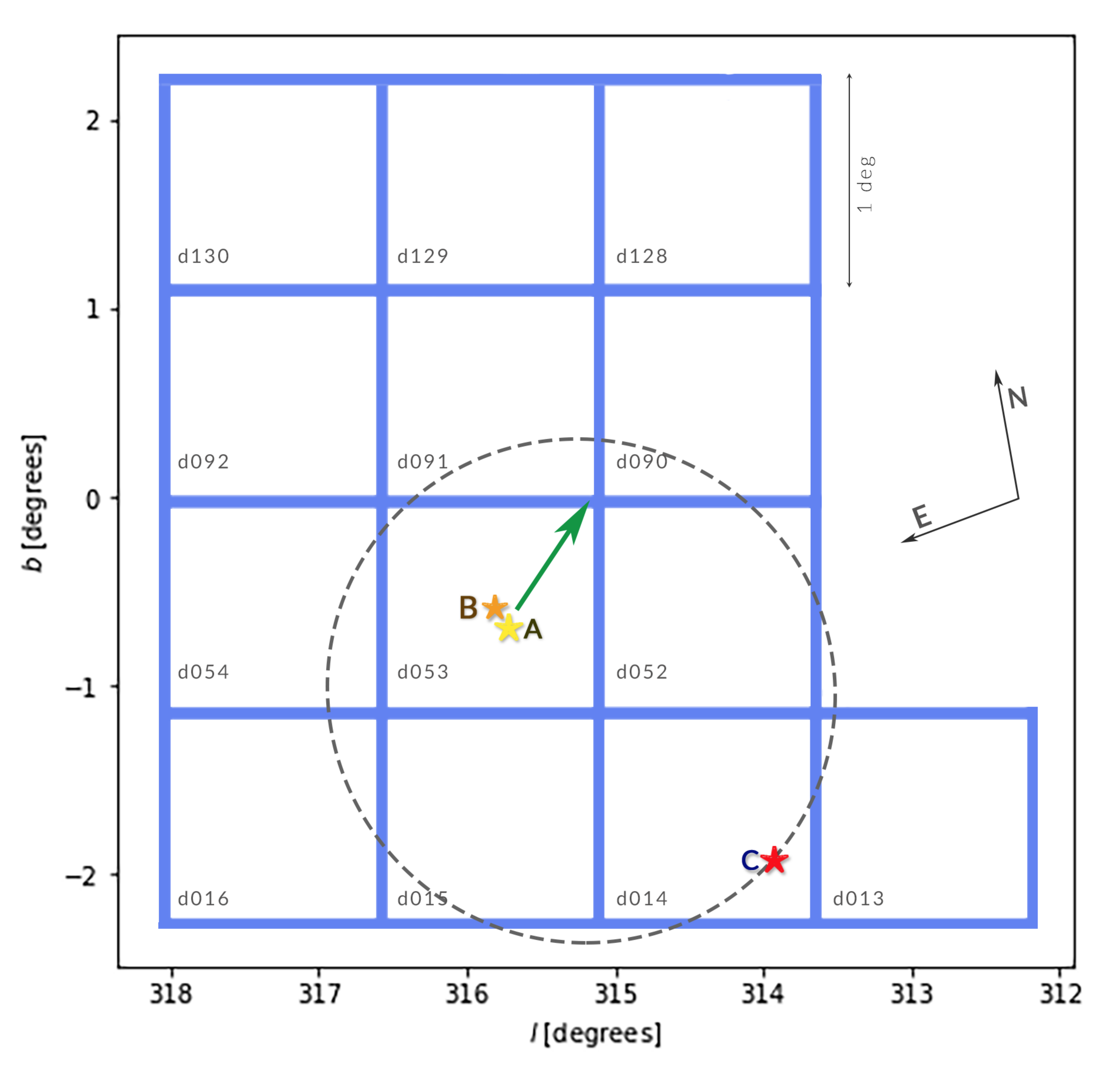}
\caption{Field of view analyzed in this study, each rectangle corresponds to one VVV tile. The position of the $\alpha$ Cen AB system is marked in yellow-orange (separation not to scale) stars and Proxima as a red star. The grey ellipse shows an approximate orbit of Proxima around the AB pair, from \citet{Kervella2017}. The green arrow shows the proper motion of the system. The total area covered by this search is $\sim$19 sq. degrees. The limiting distance of 7\,000 AU (and 1\,200 to Proxima) is given by the distance to the South-East limit of the tile 'd015' (and 'd014' for Proxima). To the North-East and North-West the limiting separation almost reach 20.000 AU from $\alpha$ Cen AB.}
\label{fig:area_surveyed}
\end{figure}

\section{Methods}
\subsection{Visual inspection of images}
\begin{figure}
\centering
   \includegraphics[width=\linewidth]{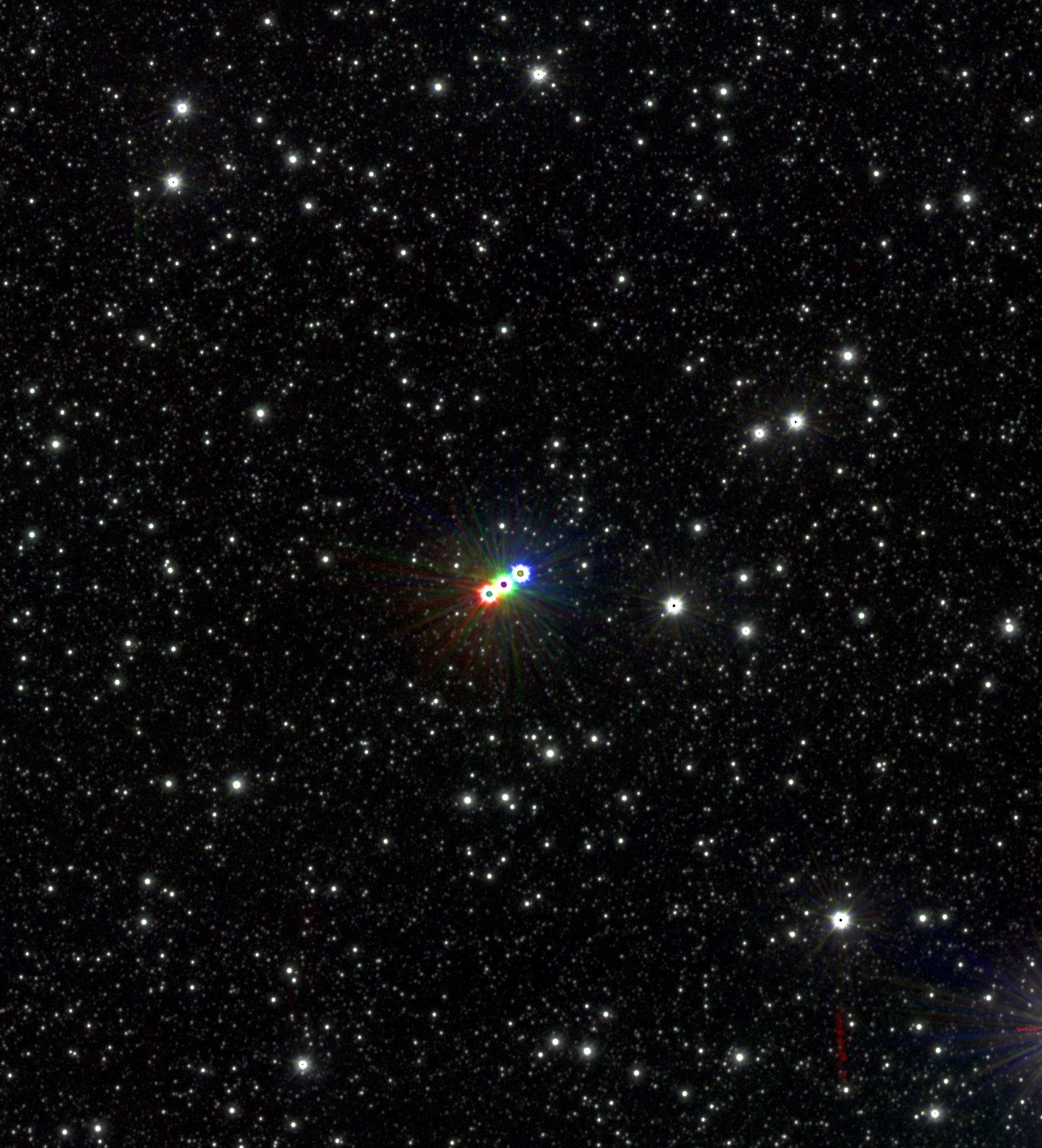}
\caption{Colour composite image of Proxima, we used three $K_S$ epochs, 2010, 2012 and 2014 colours are red green and blue respectively. The motion is evident on the image at the bottom right. This image highlight the motion of the bright star Proxima, but additional colour enhancement was used to search for the faintest moving target across the images.}
\label{fig:hpm_source}
\end{figure}

We created false colour images for 13 tiles using $K_S$ images taken at 3 epochs separated in time by approximately 2 years from each other. The total area covered by the images was $\sim$19\,deg$^{2}$ (See Fig. \ref{fig:area_surveyed}). 
 A source with the same motion of the $\alpha$ Cen system would have left an easily recognizable colour trace with the same position angle as the proper motion of the $\alpha$ Cen system. All non-variable sources would appear white, variables of high amplitude would have a point-like shape and the colour skewed to the colour assigned to the epoch at maximum brightness. Very high proper motion sources like solar system objects would only be detected in one image and be detected as a point-like source with only one colour. Other high proper motions sources would appeared as elongated sources with red and blue colours at the edges (Fig. \ref{fig:hpm_source} shows this feature for Proxima). Other artifacts like ghosts, diffraction spikes etc. would not produce a linear trace of point-like sources in any case, so it did not affect our visual search.
 
With this method we can reject the presence of an extra component to the system down to $K_S\sim$17-17.5 mag, which is the 5$\sigma$ point source magnitude limit detection per epoch, and mainly discard sources around very bright and saturated stars. 

\subsection{Source catalog cross matching}
\label{Sect:auto}
Following the visual inspection, we retrieved catalogs for 13 different pointings (tiles) from two epochs separated by $\sim$5 years in the $Y$ and $J$ band from the VVV survey. 
We choose to use the images in $Y$ and $J$ band because these images are deeper than $H$ and $K_S$ and also more sensitive to ultra cool ($\rm T_{eff}$ $\leq$500\,K) brown dwarfs (BD) in the NIR , than the $K_S$ band \citep{Morley2014, Beamin2014,Luhman2016,Zapatero2016,Schneider2016,Leggett2017}. $H$ band is also sensitive to UCDs, but in our survey is shallower than $Y$ and $J$ bands, so this would not improve the detectability of a UCD.

The total time span between the two $Y$ and $J$ band observations is 5 years (2010-2015). Additionally, $Y$ and $J$ band epochs are usually not taken simultaneously. Images from the same year for the $Y$ and $J$ bands for 8 out of the 13 tiles were obtained with a time difference larger than 20 days at least in one epoch, 20 days is the required time for a source co-moving with the $\alpha$ Cen system to move $\sim$ 0.6 pixels in the VIRCAM camera, and hence produce a shift in the centroid of 0.1'' of the  background source, assuming both sources have similar fluxes. This implies that we effectively had 3 or 4 epochs, decreasing significantly the already low chance of an alignment between a possible companion of $\alpha$ Cen system and a background source. The dates of each individual image in $Y$ and $J$ bands are given in the Table \ref{Tab:obs} in the appendix.

The first epoch was observed usually between March and April 2010, and the second epoch around May-June 2015.
For these catalogs the 5$\sigma$ limiting magnitudes are $Y\sim$19.3 and $J\sim$19.0 mag. 
A list of the values per tile is given in Table \ref{Tab:limits}.

\begin{figure}
\centering
   \includegraphics[scale=0.36]{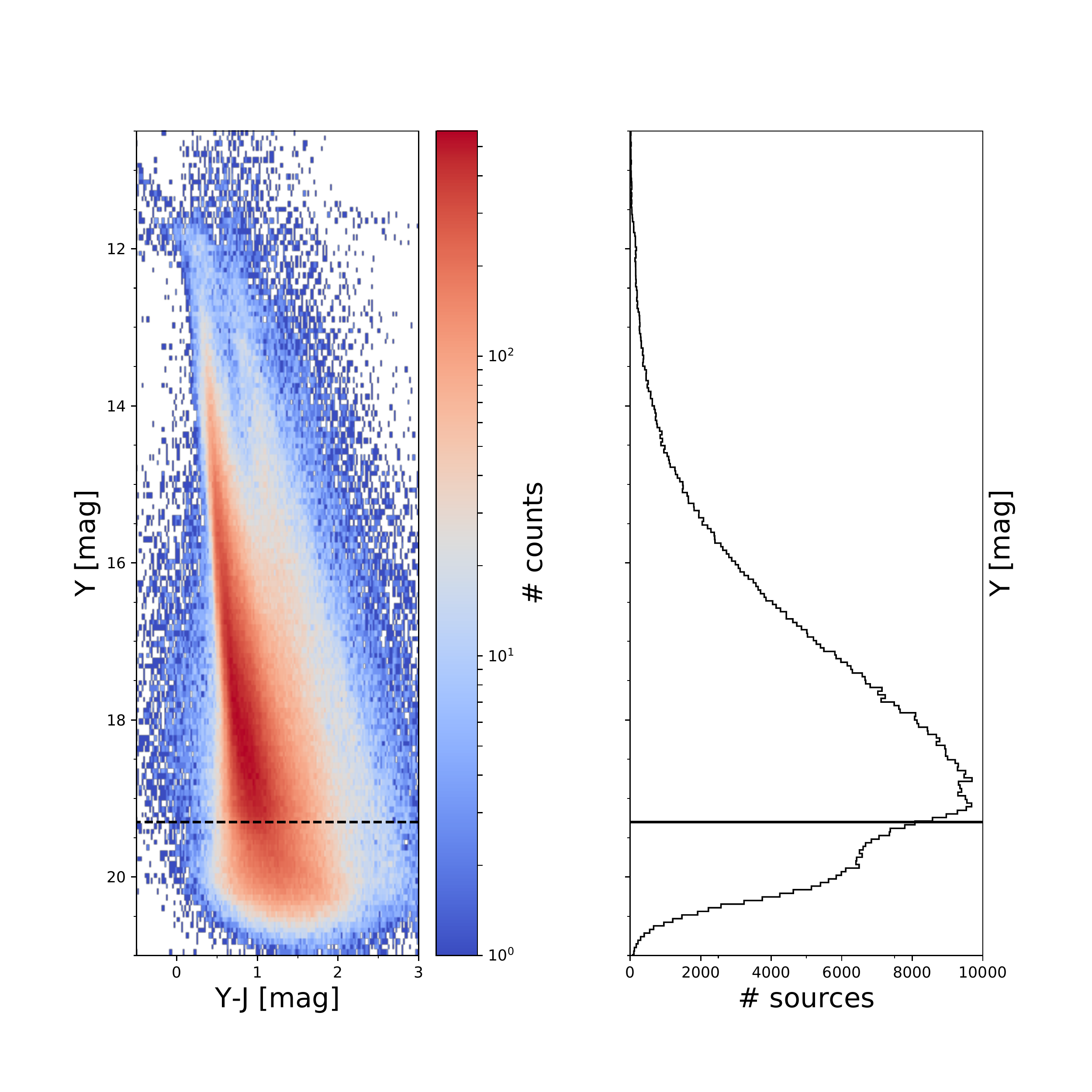}
      \caption{Left panel: Colour magnitude diagram of Tile d053,  right panel: histogram of sources in the $Y$ band presented in the CMD }
         \label{fig:cmd}
 \end{figure}

A typical colour magnitude diagram in the $Y$ and $J$ bands is shown in Fig. \ref{fig:cmd}, in this case we selected sources from Tile d053 (containing $\alpha$ Cen) which is the least detection-favoring tile due to the saturation, spikes and ``image ghosts''. Nevertheless, over 700,000 sources were cross-matched between the two bands. Additionally, we included the histogram in $Y$ magnitude and the 5$\sigma$  photometric detection limit ($Y$=19.3 mag) as a dashed line in the colour magnitude diagram (CMD) and solid black line and in the histogram. The remaining 12 tiles analyzed in this study share similar number counts and overall shape of the CMD, with a very strong disk sequence  and a less populated  sequence of giant stars to the right, typical  for the inner region of the disk population and the expected colour spread due to interstellar extinction.

\begin{figure}
\centering
   \includegraphics[scale=0.55]{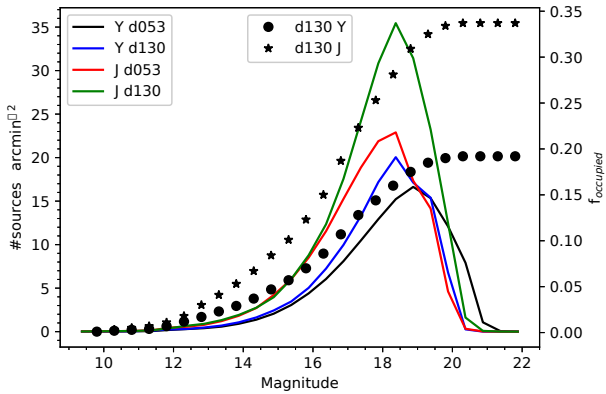}
      \caption{The density of sources per 0.5 magnitude bin per square arcminute for two tiles d130, the one with the largest number of sources, and d053, the tile that contains $\alpha$ Cen AB, both are the worst case scenarios in terms of highest source density and artifacts generated by the presence of an extremely bright source. Additionally the right axis display the cumulative area covered by stars brighter at any given magnitude, only tile d130 is shown, all the other tiles display a lower fraction of the image covered by stars} 
         \label{fig:dens}
 \end{figure}  

\begin{table}
\caption{5$\sigma$ limiting magnitude for the Tiles analyzed in this work.}            
\label{Tab:limits}     
\centering                         
\begin{tabular}{c@{    }  c@{       } c@{         }  c@{         } c@{       } }
\hline\hline                 
Tile name  &  Y$_{2010}$  &  Y$_{2015}$ & J$_{2010}$ & J$_{2015}$ \  \\
& [mag] & [mag] & [mag] & [mag]\\
\hline             
d013 & 19.55 &  19.34 & 19.24 & 19.14 \\ 
d014 & 19.50 &  19.30 & 18.71 & 19.05 \\ 
d015 & 19.48 &  19.44 & 18.92 & 18.92 \\ 
d016 & 19.44 &  19.41 & 18.77 & 19.03 \\ 
d052 & 19.86 &  19.54 & 19.57 & 19.33 \\ 
d053 & 19.97 &  19.61 & 19.29 & 19.17 \\ 
d054 & 20.13 &  19.69 & 19.43 & 19.26 \\ 
d090 & 20.01 &  19.72 & 19.53 & 19.4 \\ 
d091 & 19.91 &  19.57 & 19.55 & 19.45 \\ 
d092 & 19.60 &  19.56 & 19.36 & 19.28 \\ 
d128 & 19.85 &  19.64 & 19.42 & 19.25 \\ 
d129 & 19.62 &  19.51 & 19.41 & 19.39 \\ 
d130 & 19.53 &  19.42 & 19.32 & 19.25 \\ 
\hline \\
Adopted & 19.3 & & 19.& \\
\hline                                   
\end{tabular}
\end{table}

To search for co-moving companions we first performed a cross-match between the two epochs of the same band ($Y$ and $J$ bands respectively), we used STILTS to perform the cross-match \citep{Taylor2005}. A 0.7\arcsec\ tolerance radius (2 pixels) was defined for the match, and we kept only sources that do not have a counterpart in the other epoch, effectively removing the low proper motion sources.

To the remaining stars, we applied the proper motion and parallax motion corresponding to each member of the $\alpha$ Cen system separately ($\alpha$ Cen AB barycentric motion from \citet{Kervella2016a} and Proxima from \citet{Benedict1999}, to the catalog from year 2010 and performed a new cross-match with the remaining sources in the 2015 catalog (we used the python jplephem\footnote{\url{https://pypi.python.org/pypi/jplephem}} software to calculate the parallax factors at each epoch). The values of proper motion and parallax used to shift the catalogs are available in Table \ref{Tab:astrometry_alfa} \footnote{Repeating  the process with the older values from Hipparcos astrometry from \citealt{vanleeuwen2007} and newer astrometry for Proxima  from \citealt{Lurie2014} does not give positive results either}. 

\begin{table}
\caption{Astrometry of $\alpha$ Cen system.}            
\label{Tab:astrometry_alfa}     
\centering                         
\begin{tabular}{c@{       }  c@{          } c@{       } c@{     } }
\hline\hline                 
 Star name  &  $\mu \alpha$cos\,$\delta$ &  $\mu \delta$ & $\pi$\\
            & mas\,yr$^{-1}$ & mas\,yr$^{-1}$& mas  \\
\hline             
$\alpha$ Cen A$^a$    & -3619.9 & 693.8 & 747.17     \\
$\alpha$ Cen B$^a$    & -3619.9 & 693.8 & 747.17    \\
$\alpha$ Cen C$^b$ (Pr\'oxima)    & -3773.84 & 770.54 & 768.7 \\

\hline     
\end{tabular}
\\
a. \citet{Kervella2016a}, b. \citet{Benedict1999}
\end{table}

For the cross-match between the catalogs, we used a 0.35\arcsec\ tolerance radius and a allowed a difference of up to 0.3 magnitudes for the corresponding band, which is the photometric uncertainty in the $J$ band at the 5$\sigma$ detection limit. Given the high density of sources towards the galactic inner disk ($\sim 900\,000$ objects per sq. degree for the VVV limiting magnitude of $J\sim$19.5) after the cross-match we obtained nearly 50 sources per tile per band, per $\alpha$ Cen stellar member. 

A match between these co-moving candidates sources in the $Y$ and $J$ bands was performed, but not a single object was detected in the two bands, which was expected if there were no additional companions.

Nevertheless we explore if there might be one real  source detected in a single band.
Most of the sources that passed the first cut were flagged as noise detections. So we decided to apply one more filter, selecting only sources that are flagged as stellar in the CASU catalogs (flag for stellar objects is -1),  the last criterion rejected most sources around spikes of saturated stars, or near the edge of the detectors, and blended objects. The remaining candidates per tile, per band and per stellar member of the $\alpha$ Cen system now were reduced to five. We did a visual inspection of these sources in 1\arcmin\ $\times$ 1\arcmin\ images,  and searched for the source in the original  $Y$ and $J$ band images simultaneously \footnote{On tile d129 we could only use $J$ band because the $Y$ band image taken in 2010 was defective, producing twice the detections all over the field of view.}.  
We eliminated all these candidates, because in at least one of the bands we could see  faint sources in both positions, but one of them not being detected by the finding/photometry routine,  because of sensitivity or contrast issues caused by artifacts. For a handful of objects where we had doubts in both the $Y$ and $J$ bands we looked at $Z$, $H$ and $K_S$ band images, and we were able to confirm that there was a background source in each expected position.
We did not find any source that passes all the matching criteria and the final visual inspection check.

We considered the possibility that at the epoch of observation a faint source might remain undetected because it was projected on the same position of a brighter background star. In Fig. \ref{fig:dens} the number of sources per square arc minute per 0.5\,magnitude bin is plotted, and also the cumulative fraction of the pixels on the image occupied by sources brighter at each given magnitude.  We only show here the information for tiles d053 and d130 which are the worst case scenario, containing $\alpha$ Cen AB and the highest number of sources in the $J$ band respectively. It can be seen that below $J$,$Y\sim$16 mag there is less than 10$\%$ chance of alignment, but it reaches to almost 30$\%$ at the limiting magnitude in $J$ band and 19$\%$ at the $Y$ band limit. Thus, the completeness of this search is at least 80$\%$, considering the results in the $Y$ band. Nevertheless, all the fields were observed at different times in $Y$ and $J$ band as explained above, which will decrease the effective area of occupied pixels and hence increase the completeness.
Simply multiplying the covered fraction of the independent images would be the easiest way to calculate the final covering fraction, but most of the brightest sources will be detected over the same pixels, and introduce a strong correlation, therefore the simple multiplication would be over optimistic, we adopt then a conservative 85-88$\%$ of completeness for the worst case scenario of tile d130, and $\gtrsim$90$\%$ for the remaining tiles

\section{Discussion }
\label{sec:discussion}
For the $\alpha$ Cen system, there is a plethora of previous studies, and several properties are well constrained, a parallax of 747.17$\pm$ 0.60 mas (1.338$\pm$0.001\,pc) \citep{Kervella2016a},
a metallicity ([Fe/H]) 0.23$\pm$ 0.05 dex \citep{Ramirez2013}, and an  age range between 4 and 7 Gyr \citep{Eggenberger2004,Mamajek2008,Boyajian2013,Bazot2016}.
The most stringent constraint on the presence of an extra companion in our study comes from the $Y$ and $J$ photometry. No source is detected up to a magnitude 19.3 and 19.0 mag in $Y$ and $J$ respectively, within separations up to  of 7\,000\,AU from $\alpha$ Cen AB system. The same photometric limits apply for Proxima, no companions to Proxima were found up to separations of 1\,200 AU. 
To transform these magnitude limits to physical parameters, i.e. effective temperatures and masses, we used different atmospheric and evolutionary models.

\begin{figure}
\centering
   \includegraphics[scale=0.51]{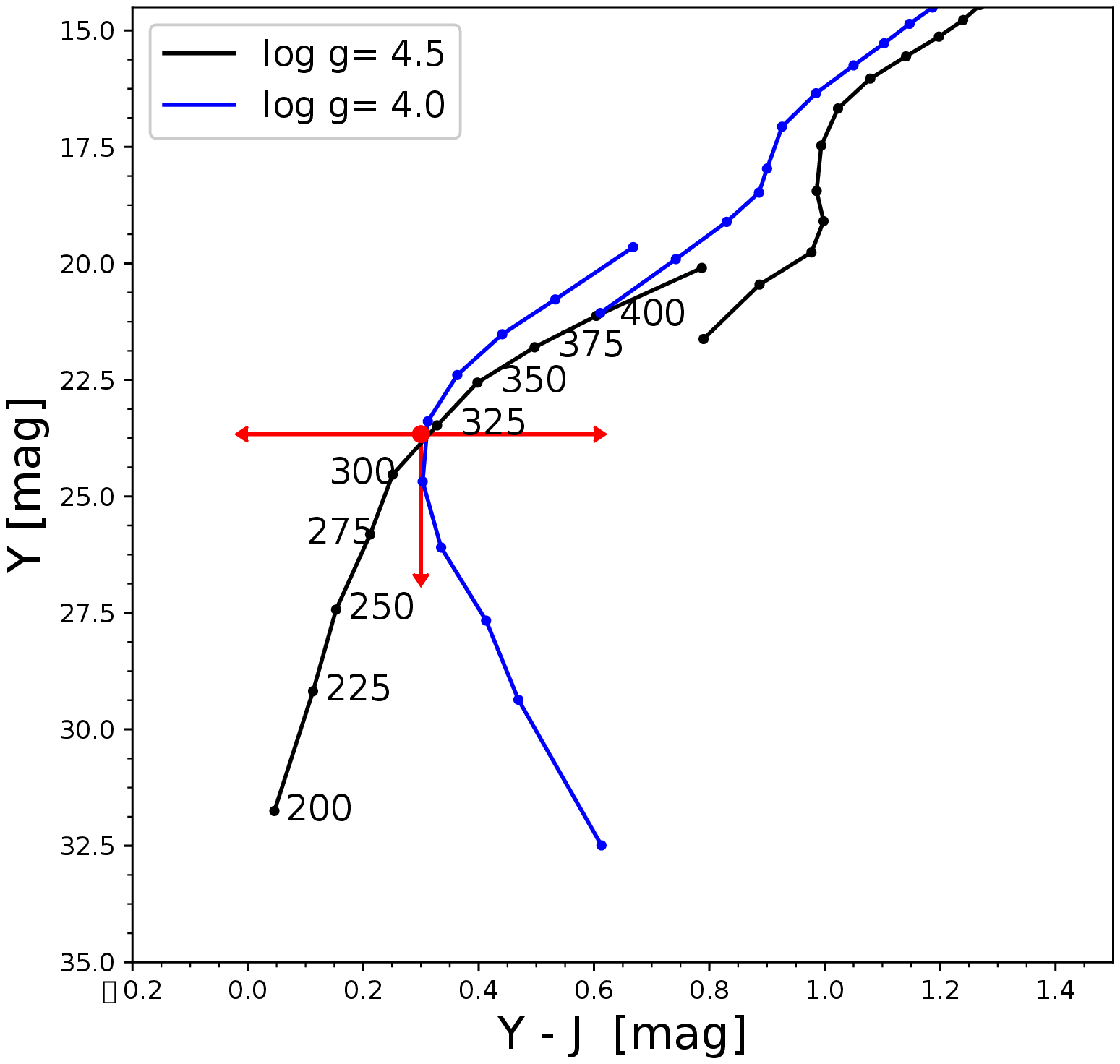}
      \caption{Colour and absolute magnitude from the atmospheric models from \citealt{Morley2012} for T and Y dwarfs and ,\citealt{Morley2014} for Y dwarfs.
      For both models we assumed a sedimentation factor of 5 and varied $log (g)$ between  4 and 4.5. The ``jump''  between the models at $\rm T_{eff}$ $\sim 400K$ can be explained by the different cloud treatment in the two models.
The red dot indicates the limiting magnitude in $Y$ band and the colour given by the limits in $Y$ and $J$ bands.}
         \label{fig:morley14}
 \end{figure}

First we used the BD cloudy models from  \cite{Morley2012} and \cite{Morley2014}\footnote{The models from \cite{Morley2012,Morley2014} and \cite{Saumon2012} were taken from \url{http://www.ucolick.org/~cmorley/cmorley/Models.html}}. The first one only reaches up to $\rm T_{ eff}$=400\,K and considers  Na$_2$S, MnS, ZnS, Cr, KCl condensate clouds. The latter models assumes a 50 $\%$ cloud covered atmosphere, composed of H$_2$O ice in addition to the Na$_2$S, KCl, ZnS, MnS, and Cra, These models reach $\rm T_{ eff}$=200\,K .
Based  on these models we were able to discard objects with  $\rm T_{eff}$ $>$325\,K for any given combination of sedimentation factor and surface gravity, these results can be seen in Fig \ref{fig:morley14}, where we plot $Y-J$ color, against absolute $Y$ magnitude.

\begin{figure}
\centering
   \includegraphics[scale=0.51]{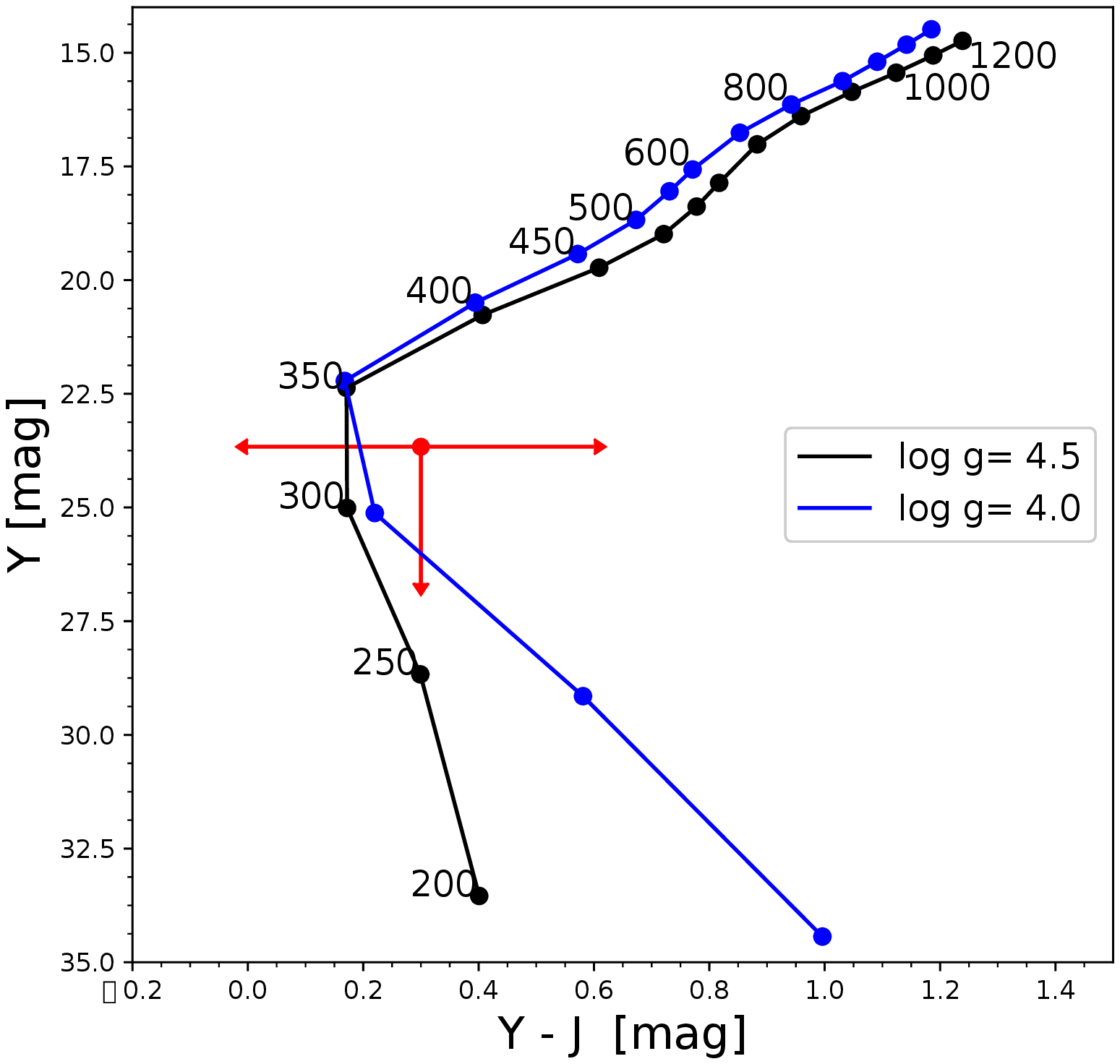}
   \includegraphics[scale=0.51]{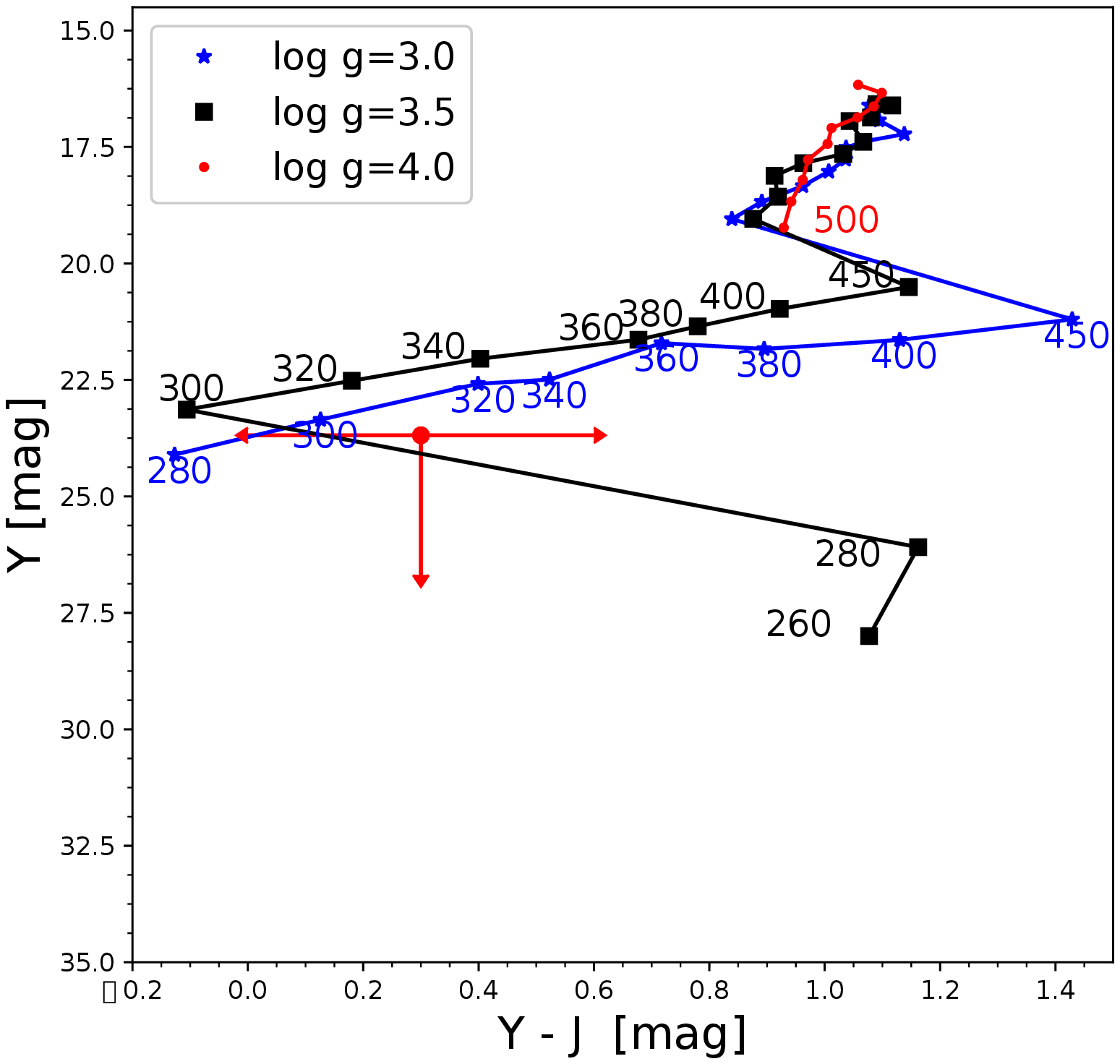}
      \caption{Colour and magnitude from the cloud-free atmospheric models upper panel: \citet{Saumon2012} for T and Y dwarfs. Assuming $log (g)$  between 4 and 4.5. 
The red dot and arrows indicates our upper limit as in figure 5.
Bottom panel: BT-Settl2011 models \citep{Allard2012} for late T and Y dwarfs }
\label{fig:mods12}
 \end{figure}

Second, we tested with the atmospheric models of \cite{Saumon2012}. These models include an improved line list of the NH$_3$ molecule, and of the collision-induced absorption of molecular hydrogen (H$_2$), no clouds opacities were considered for these temperatures. The colours were calculated using the \citealt{Saumon2008} cloud-free evolution model grids. 
The limits for this model are shown in the upper panel of Fig. \ref{fig:mods12}.
Doing a simple linear interpolation of the data in the $Y$ and $J$ bands against effective temperature, considering values of $log (g)$ between 4 and 4.5, and evaluating the limits obtained in $Y$ and $J$ bands (19.3 and 19.0 mag respectively) we derived upper limits for the effective temperature of 326\,K and 322\,K, for the $Y$ and $J$ bands respectively.

Finally, we tested the BT-Settl models \citep{Allard2012}\footnote{ \url{https://phoenix.ens-lyon.fr/Grids/BT-Settl/CIFIST2011/COLORS/}}, which uses the updated solar abundances from  \cite{Caffau2011}, also account for a calibration of the mixing length based on Radiation Hydrodynamics  simulations by \cite{Freytag2010} and adjustments to the MLT equations.
To transform between the absolute magnitudes given at the surface of the BD provided by models, we assumed a radius of  0.1 R$_\odot$ which is the expected for these kind of objects at ages around 4-7 Gyr \citep{Burrows2001}.
Unfortunately, only objects with low surface gravities are available in the public grid of models at these low temperatures ($log (g)$ 3.0 and 3.5), for higher gravities ($log (g)$ 4.0 and 4.5) models are available for effective temperatures above 500\,K.
In the bottom panel of Fig. \ref{fig:mods12} it can be seen that the models with higher gravity and $\rm T_{ eff}$>500\,K are far above from our limit, and indeed an extension of the models to higher gravities is required to compare the temperature limits with the other set  of models. We also plot the lower gravity models to roughly estimate a limit in $\rm T_{ eff}$, and we can clearly see that objects with temperatures below 300\,K are ruled out.  


Assuming the distance from \citet{Kervella2016a}, and using the evolutionary models of
 \citet{Saumon2008}, with the atmosphere calculation described in \citet{Saumon2008} and \citet{Marley2002}. The cloudless model at the $\sim$320\,K limit implies a mass of 10.5\,M$_{Jup}$, 12.6\,M$_{Jup}$, and 14.7\,M$_{Jup}$,
for 4, 6 and 8 Gyr and 9.4\,M$_{Jup}$, 10.5\,M$_{Jup}$, 12.6\,M$_{Jup}$, for the cloudy model (with cloud sedimentation factor 2), respectively. All these calculations were made assuming solar metallicity.

Our magnitude limits in $J$ are the same as those of  \citet{Mesa2017} for separations below 0.5 AU. Here we calculated a higher mass limit than \citet{Mesa2017} due to the use of an updated set of atmospheric models. If the same set of models is applied to their dataset, the estimated limiting masses of the planets would increase by 2-3 M$_{Jup}$.

\section{Conclusions} 
\label{conclusions}

We have carried out a deep and wide search for other members of the $\alpha$ Cen system using the VVV near-IR images. No additional companions were found around the $\alpha$ Cen AB system.
In total we explored a $\sim$19 sq. degree region around the $\alpha$ Cen AB system, ranging up to 7\,000 AU to the south east direction and nearly 20\,000 AU to the North-East and North-West direction. Also, no companions were found around Proxima within 1\,200 AU.

Our search considered a visual inspection in the $K_S$ band and using photometric 5$\sigma$ limits in the $Y$, $J$ bands. The final limit excludes the presence of a BD/planet with a mass above 9.5-14.5 M$_{Jup}$, model and age dependent.
Our search extended the limits on possible co-moving companions to the $\alpha$ Cen AB system and also Proxima, to  greater distances than previous attempts, complementing previous studies using radial velocities \citep{Dumusque2012,anglada-escude2016}, higher spatial resolution imaging \citep{Kervella2006a,Mesa2017}, deep optical imaging \citep{Kervella2007a} and astrometric searches \citep{Benedict1999}.

An extended  search will be possible in the following 2-3 years making use of the on-going VVV extended survey (VVVX), which will extend the observed area 2.2 degrees more in galactic latitude (positive and negative), this will allow us to place limits up  to separations of $\sim$18\,000 AU in every direction.

\section*{Acknowledgements}
J.C.B. Acknowledge support from programa ESO-C\'omite mixto gobierno de Chile. D.M. acknowledges project support from Basal
Center for Astrophysics and Associated Technologies CATA PFB-06 and Fondecyt grant No. 1170121.  Support for
D.M, and R.K is provided by the Ministry of Economy,
Development, and Tourism's Millennium Science Initiative through grant
IC120009, awarded to The Millennium Institute of Astrophysics, MAS.
 A. Bayo acknowledges financial support from the Proyecto Fondecyt Iniciaci\'on 11140572.
The authors acknowledge the work by the referee, Pierre Kervella, for his very useful comments and suggestions which helped to improve this manuscript.
This research has made use of the SIMBAD database, operated at CDS, Strasbourg, France
This research made use of Astropy, a community-developed core Python package for Astronomy \cite{astropy2013}.





\begin{thebibliography}{99}
\bibitem[Allard et al.(2012)]{Allard2012} Allard, F., Homeier, D., \& Freytag, B.\ 2012, Royal Society of London Philosophical Transactions Series A, 370, 2765
\bibitem[Anglada-Escud{\'e} et al.(2016)]{anglada-escude2016} Anglada-Escud{\'e}, G., Amado, P.~J., Barnes, J., et al.\ 2016, \nat, 536, 437 
\bibitem[Astropy Collaboration et al.(2013)]{astropy2013} Astropy Collaboration, Robitaille, T.~P., Tollerud, E.~J., et al.\ 2013, \aap, 558, A33 

\bibitem[Baraffe et al.(2002)]{Baraffe2002} Baraffe, I., Chabrier, G., Allard, F., \& Hauschildt, P.~H.\ 2002, \aap, 382, 563 
\bibitem[Baraffe et al.(2003)]{Baraffe2003} Baraffe, I., Chabrier, G., Barman, T.~S., Allard, F., \& Hauschildt, P.~H.\ 2003, \aap, 402, 701 
\bibitem[Bazot et al.(2016)]{Bazot2016} Bazot, M., Christensen-Dalsgaard, J., Gizon, L., \& Benomar, O.\ 2016, \mnras, 460, 1254 

\bibitem[Beam{\'{\i}}n et al.(2013)]{Beamin2013} Beam{\'{\i}}n, J.~C., Minniti, D., Gromadzki, M., et al.\ 2013, \aap, 557, LL8
\bibitem[Beam{\'{\i}}n et al.(2014)]{Beamin2014} Beam{\'{\i}}n, J.~C., Ivanov, V.~D., Bayo, A., et al.\ 2014, \aap, 570, L8 

\bibitem[Beam{\'{\i}}n et al.(2015)]{Beamin2015} Beam{\'{\i}}n, 
J.~C., Ivanov, V.~D., Minniti, D., et al.\ 2015, \mnras, 454, 4054
\bibitem[Beamin et al.(2017)]{Beamin2017} Beamin, J.~C., Mendez, R.~A., Smart, R.~L., et al.\ 2017, arXiv:1703.05817 
\bibitem[Bendek et al.(2015)]{Bendek2015} Bendek, E.~A., Belikov, R., Lozi, J., et al.\ 2015, \procspie, 9605, 960516 
\bibitem[Benedict et al.(1999)]{Benedict1999} Benedict, G.~F., McArthur, B., Chappell, D.~W., et al.\ 1999, \aj, 118, 1086 
\bibitem[Boyajian et al.(2013)]{Boyajian2013} Boyajian, T.~S., von Braun, K., van Belle, G., et al.\ 2013, \apj, 771, 40 
\bibitem[Burrows et al.(2001)]{Burrows2001} Burrows, A., Hubbard, W.~B., Lunine, J.~I., \& Liebert, J.\ 2001, Reviews of Modern Physics, 73, 719 

\bibitem[Caffau et al.(2011)]{Caffau2011} Caffau, E., Ludwig, H.-G., Steffen, M., Freytag, B., \& Bonifacio, P.\ 2011, \solphys, 268, 255 
\bibitem[Dalton et al.(2006)]{Dalton2006} Dalton, G.~B., Caldwell, M., Ward, A.~K., et al.\ 2006, \procspie, 6269, 62690X
\bibitem[D{\'e}k{\'a}ny et al.(2013)]{Dekany2013} D{\'e}k{\'a}ny, I., Minniti, D., Catelan, M., et al.\ 2013, \apjl, 776, LL19
\bibitem[Demory et al.(2015)]{Demory2015} Demory, B.-O., Ehrenreich, D., Queloz, D., et al.\ 2015, \mnras, 450, 2043 

\bibitem[Dumusque et al.(2012)]{Dumusque2012} Dumusque, X., Pepe, F., Lovis, C., et al.\ 2012, \nat, 491, 207 
\bibitem[Dupuy \& Liu(2012)]{Dupuy2012} Dupuy, T.~J., \& Liu, M.~C.\ 2012, \apjs, 201, 19 
\bibitem[Eggenberger et al.(2004)]{Eggenberger2004} Eggenberger, P., Charbonnel, C., Talon, S., et al.\ 2004, \aap, 417, 235 


\bibitem[Emerson \& Sutherland(2010)]{Emerson2010} Emerson, J., \& Sutherland, W.\ 2010, The Messenger, 139, 2 
\bibitem[Endl and K\"{u}rster(2008)]{Endl2008} Endl, M. \& K\"{u}rster, M.\ 2008, \aap, 488, 1149
\bibitem[Freytag et al.(2010)]{Freytag2010} Freytag, B., Allard, F., Ludwig, H.-G., Homeier, D., \& Steffen, M.\ 2010, \aap, 513, A19 
\bibitem[Gonzalez et al.(2011)]{Gonzalez2011} Gonzalez, O.~A., Rejkuba, M., Minniti, D., et al.\ 2011, \aap, 534, LL14
\bibitem[Hempel et al.(2014)]{Hempel2014} Hempel, M., Minniti, D., D{\'e}k{\'a}ny, I., et al.\ 2014, The Messenger, 155, 29 
\bibitem[Ivanov et al.(2013)]{Ivanov2013} Ivanov, V.~D., Minniti, D., Hempel, M., et al.\ 2013, \aap, 560, AA21
\bibitem[Ivanov et al.(2015)]{Ivanov2015} Ivanov, V.~D., Vaisanen, P., Kniazev, A.~Y., et al.\ 2015, \aap, 574, A64 
\bibitem[Kervella et al.(2017)]{Kervella2017} Kervella, P., Th{\'e}venin, F., \& Lovis, C.\ 2017, \aap, 598, L7 
\bibitem[Kervella et al.(2016)]{Kervella2016a} Kervella, P., Mignard, F., M{\'e}rand, A., \& Th{\'e}venin, F.\ 2016, \aap, 594, A107 
\bibitem[Kervella et al.(2016)]{Kervella2016b} Kervella, P., Th{\'e}venin, F., \& Lovis, C.\ 2016, arXiv:1611.03495 
\bibitem[Kervella \& Th{\'e}venin(2007)]{Kervella2007a} Kervella, P., \& Th{\'e}venin, F.\ 2007, \aap, 464, 373 
\bibitem[Kervella et al.(2006)]{Kervella2006a} Kervella, P., Th{\'e}venin, F., Coud{\'e} du Foresto, V., \& Mignard, F.\ 2006, \aap, 459, 669 

\bibitem[Kurtev et al.(2017)]{Kurtev2017} Kurtev, R., Gromadzki, M., Beam{\'{\i}}n, J.~C., et al.\ 2017, \mnras, 464, 1247 
\bibitem[Leggett et al.(2017)]{Leggett2017} Leggett, S.~K., Tremblin, P., Esplin, T.~L., Luhman, K.~L., \& Morley, C.~V.\ 2017, arXiv:1704.03573 
\bibitem[Luhman \& Esplin(2016)]{Luhman2016} Luhman, K.~L., \& Esplin, T.~L.\ 2016, \aj, 152, 78 
\bibitem[Lurie et al.(2014)]{Lurie2014} Lurie, J.~C., Henry, T.~J., Jao, W.-C., et al.\ 2014, \aj, 148, 91 
\bibitem[Mamajek \& Hillenbrand(2008)]{Mamajek2008} Mamajek, E.~E., \& Hillenbrand, L.~A.\ 2008, \apj, 687, 1264-1293 
\bibitem[Minniti et al.(2010)]{Minniti2010} Minniti, D., Lucas, P.~W., Emerson, J.~P., et al.\ 2010, \na, 15, 433 
\bibitem[Minniti et al.(2014)]{Minniti2014} Minniti, D., Saito, R.~K., Gonzalez, O.~A., et al.\ 2014, \aap, 571, AA91
\bibitem[Marley et al.(2002)]{Marley2002} Marley, M.~S., Seager, S., Saumon, D., et al.\ 2002, \apj, 568, 335 
\bibitem[Morley et al.(2012)]{Morley2012} Morley, C.~V., Fortney, J.~J., Marley, M.~S., et al.\ 2012, \apj, 756, 172 
\bibitem[Morley et al.(2014)]{Morley2014} Morley, C.~V., Marley, M.~S., Fortney, J.~J., et al.\ 2014, \apj, 787, 78 
\bibitem[Mesa et al.(2017)]{Mesa2017} Mesa, D., Zurlo, A., Milli, J., et al.\ 2017, \mnras, 466, L118 

\bibitem[Pourbaix \& Boffin(2016)]{Pourbaix2016} Pourbaix, D., \& Boffin, H.~M.~J.\ 2016, \aap, 586, A90 
\bibitem[Quarles \& Lissauer(2016)]{Quarles2016} Quarles, B., \& Lissauer, J.~J.\ 2016, \aj, 151, 111 
\bibitem[Rajpaul et al.(2016)]{Rajpaul2016} Rajpaul, V., Aigrain, S., \& Roberts, S.\ 2016, \mnras, 456, L6 
\bibitem[Ram{\'{\i}}rez et al.(2013)]{Ramirez2013} Ram{\'{\i}}rez, I., Allende Prieto, C., \& Lambert, D.~L.\ 2013, \apj, 764, 78 
\bibitem[Saito et al.(2012)]{Saito2012} Saito, R.~K., Hempel, M., Minniti, D., et al.\ 2012, \aap, 537, A107 
\bibitem[Saumon \& Marley(2008)]{Saumon2008} Saumon, D., \& Marley, M.~S.\ 2008, \apj, 689, 1327-1344 
\bibitem[Saumon et al.(2012)]{Saumon2012} Saumon, D., Marley, M.~S., Abel, M., Frommhold, L., \& Freedman, R.~S.\ 2012, \apj, 750, 74 
\bibitem[Schneider et al.(2016)]{Schneider2016} Schneider, A.~C., Cushing, M.~C., Kirkpatrick, J.~D., \& Gelino, C.~R.\ 2016, \apjl, 823, L35 
\bibitem[Sirbu et al.(2017)]{Sirbu2017} Sirbu, D., Thomas, S., \& Belikov, R.\ 2017, arXiv:1704.05441 

\bibitem[Smith et al.(2015)]{Smith2015} Smith, L.~C., Lucas, P.~W., Contreras Pe{\~n}a, C., et al.\ 2015, \mnras, 454, 4476
\bibitem[Smith et al.(2017 subm.)]{Smith2017} Smith, L.~C. subm

\bibitem[Taylor(2005)]{Taylor2005} Taylor, M.~B.\ 2005, Astronomical Data Analysis Software and Systems XIV, 347, 29
\bibitem[Thomas et al.(2015)]{Thomas2015} Thomas, S., Belikov, R., \& Bendek, E.\ 2015, \apj, 810, 81 

\bibitem[van Leeuwen(2007)]{vanleeuwen2007} van Leeuwen, F.\ 2007, \aap, 474, 653 
\bibitem[Wright et al.(2010)]{Wright2010} Wright, E.~L., Eisenhardt, P.~R.~M., \& Mainzer, A.~K. et al. 2010, \aj, 140, 1868 
\bibitem[Zapatero Osorio et al.(2016)]{Zapatero2016} Zapatero Osorio, M.~R., Lodieu, N., B{\'e}jar, V.~J.~S., et al.\ 2016, \aap, 592, A80 

\end{thebibliography}

\appendix

\section{Individual observation dates}
We list here all the epochs used to obtain the magnitude limits in this study for the $Y$ and $J$ bands.
\begin{table}
\caption{VVV individual tile frames used in this study.}            
\label{Tab:obs}     
\centering                         
\begin{tabular}{c@{   }  c@{     } c@{  } 
}
\hline\hline                 
Tile name  &  Filter  &  Date   \\
\hline             
d013 & $Y$ & 2010-03-27T03:25:15.8423  \\ 
d013 & $J$ & 2010-04-02T06:05:56.1283  \\ 
d013 & $Y$ & 2015-05-03T07:02:50.8657  \\ 
d013 & $J$ & 2015-06-08T05:23:57.1343  \\ 
d014 & $Y$ & 2010-03-27T03:38:32.1778  \\ 
d014 & $J$ & 2010-04-01T08:10:18.4379  \\ 
d014 & $Y$ & 2015-05-03T07:13:05.4679  \\ 
d014 & $J$ & 2015-06-08T05:50:55.0405  \\ 
d015 & $Y$ & 2010-03-28T05:07:37.9603  \\ 
d015 & $J$ & 2010-04-03T06:11:37.0379  \\ 
d015 & $Y$ & 2015-05-02T08:31:57.0208  \\ 
d015 & $J$ & 2015-05-04T07:29:17.1217  \\ 
d016 & $Y$ & 2010-03-29T03:54:26.5940  \\ 
d016 & $J$ & 2010-04-05T06:20:45.6914  \\ 
d016 & $Y$ & 2015-04-29T08:03:36.5387  \\ 
d016 & $J$ & 2015-05-09T07:28:21.9441  \\ 
d016 & $J$ & 2015-05-14T06:57:05.1496  \\ 
d052 & $Y$ & 2010-03-27T03:52:31.8889  \\ 
d052 & $J$ & 2010-04-02T07:01:43.5797  \\ 
d052 & $Y$ & 2015-05-03T07:24:07.6892  \\ 
d052 & $J$ & 2015-06-08T06:16:15.7338  \\ 
d053 & $Y$ & 2010-03-28T05:23:19.5800  \\ 
d053 & $J$ & 2010-04-03T06:36:25.6897  \\ 
d053 & $Y$ & 2015-05-02T08:41:46.9336  \\ 
d053 & $J$ & 2015-05-04T07:53:33.6876  \\ 
d054 & $Y$ & 2010-03-28T05:36:12.4921  \\ 
d054 & $J$ & 2010-04-03T07:02:25.9449  \\ 
d054 & $Y$ & 2015-05-02T08:52:42.7697  \\ 
d054 & $J$ & 2015-05-04T08:24:04.8148  \\ 
d090 & $J$ & 2010-03-27T04:43:39.0487  \\ 
d090 & $Y$ & 2010-03-28T04:39:26.0940  \\ 
d090 & $Y$ & 2015-04-29T07:41:48.8063  \\ 
d090 & $J$ & 2015-05-09T04:52:36.3383  \\ 
d091 & $J$ & 2010-03-27T05:48:09.3985  \\ 
d091 & $Y$ & 2010-06-25T03:28:45.5742  \\ 
d091 & $Y$ & 2015-05-03T07:34:18.5241  \\ 
d091 & $J$ & 2015-05-29T04:31:46.7876  \\ 
d091 & $J$ & 2015-06-24T02:50:02.5166  \\ 
d092 & $J$ & 2010-03-29T05:16:06.8899  \\ 
d092 & $Y$ & 2010-06-16T04:52:10.5778  \\ 
d092 & $Y$ & 2015-04-16T09:08:36.6166  \\ 
d092 & $Y$ & 2015-04-16T09:18:28.5562  \\ 
d092 & $J$ & 2015-06-01T04:10:11.8593  \\ 
d128 & $J$ & 2010-03-27T05:08:24.3063  \\ 
d128 & $Y$ & 2010-03-28T04:54:23.8370  \\ 
d128 & $Y$ & 2015-04-29T07:52:16.4593  \\ 
d128 & $J$ & 2015-05-09T05:29:48.4227  \\ 
d129 & $J$ & 2010-03-27T06:10:27.3904  \\ 
d129 & $Y$ & 2010-06-25T03:45:04.0264  \\ 
d129 & $Y$ & 2015-05-03T07:46:03.7577  \\ 
d129 & $J$ & 2015-06-24T03:29:36.1733  \\ 
d130 & $J$ & 2010-03-27T06:32:28.5134  \\ 
d130 & $Y$ & 2010-06-25T03:59:50.6191  \\ 
d130 & $Y$ & 2015-05-03T07:56:15.7642  \\ 
d130 & $J$ & 2015-06-24T04:03:23.2328  \\  
 
\hline                                   
\end{tabular}
\end{table}

\bsp	
\label{lastpage}
\end{document}